\documentclass{na48lp}
\newcommand{\epsi}{\mbox{$\varepsilon$}}
\newcommand{\eprim}{\mbox{$\varepsilon^{'}$}}
\newcommand{\eoe}{\mbox{${\rm Re}(\epsi'/\epsi)$}}
\newcommand{\pipic}{\mbox{$\pi^{+}\pi^{-}$}}
\newcommand{\pipin}{\mbox{$\pi^{0}\pi^{0}$}}
\newcommand{\ks}{\mbox{K$_{S}$}}
\newcommand{\kl}{\mbox{K$_{L}$}}
\newcommand{\twopic}{\mbox{$\pi^{+}\pi^{-}$}}
\newcommand{\twopi}{\mbox{$\pi\pi$}}
\newcommand{\twopin}{\mbox{$\pi^{0}\pi^{0}$}}
\newcommand{\ktopc}{\mbox{K$\rightarrow \pi^{+}\pi^{-}$}}
\newcommand{\ktopn}{\mbox{K$\rightarrow \pi^{0}\pi^{0}$}}
\newcommand{\kltopc}{\mbox{K$_{L}\rightarrow \pi^{+}\pi^{-}$}}
\newcommand{\kltopn}{\mbox{K$_{L}\rightarrow \pi^{0}\pi^{0}$}}
\newcommand{\kstopc}{\mbox{K$_{S}\rightarrow \pi^{+}\pi^{-}$}}
\newcommand{\kstopn}{\mbox{K$_{S}\rightarrow \pi^{0}\pi^{0}$}}
\newcommand{\kethree}{\mbox{K$_{L}\rightarrow \pi e \nu $}}
\newcommand{\kmuthree}{\mbox{K$_{L}\rightarrow \pi \mu \nu $}}
\newcommand{\ktothreepi}{\mbox{K$_{L}\rightarrow \pi^{0}\pi^{0}\pi^{0}$}}

\begin{document}

\title{Results on CP Violation from the NA48 experiment at CERN}

\author{Lydia ICONOMIDOU-FAYARD }

\address{LAL, Universit\'e d'Orsay, b\^at. 200, 91898 Orsay, France}

\author{ on behalf of the NA48 collaboration}
\address{NA48 is a Cagliari, Cambridge, CERN, Dubna, Edinburgh, 
Ferrara, Firenze, Mainz,
Orsay, Perugia, Pisa, Saclay, Siegen, Torino, Vienna and Warsaw collaboration}

\twocolumn[\maketitle\abstract{
In this article the current status and latest results of the
NA48 experiment at CERN are given. We present in more details the analysis
performed  for the \eoe ~measurement  
with  the combined statistics
accumulated during the 1998 and 1999 data periods. 
Reviewing the  NA48 rare decay program, we select to underline  
the  new results on the
branching ratio and the {\it a$_{V}$} factor for the decay K$_{L}\rightarrow \pi^{0}\gamma\gamma$ and the \kl$\rightarrow$\pipic\ e$^{+}$e$^{-}$ 
CP violating decay.}]

\section{Introduction}
\subsection{The NA48 experiment at CERN}
The main goal of NA48 is the precise measurement of \eoe\ in the neutral
kaon system. In order to 
 achieve an accuracy of $\sim2\times10^{-4}$ the optimization
of the detector and beams has been conceived to allow the minimisation
of the sensitivity to systematic effects. Moreover, several data
taking periods have been carried out, namely in 97, 98, 99 and 2000, to 
accumulate adequatly large statistics.
A first result was published in 1999, based on the data sample recorded
during the first year~\cite{firstpap}. A second, more precise result, 
was announced on May 2001, coming out from the combined 98 and 99 data
and also exploiting information from  the special 2000 run for 
checking purposes. The corresponding analysis
 will be presented in this article. During the summer 2001, 
NA48 records its last sample dedicated to \eoe\ measurement with a 
better beam duty cycle and new drift chambers.
\newline
The NA48 trigger 
bandwidth allows to record in parallel with the \eoe\ program, several rare 
kaon decay channels that  we will review here. Some of these
decays are interesting for testing Chiral Perturbation theory. Others
are related to CP-Violation. Finaly, several rare modes can probe extensions
of the Standard Model.

\subsection{CP Violation in neutral kaons}
A small non-conservation of the CP symmetry manifests
in the neutral kaon system through the observation of the forbidden
decay mode \kl $\rightarrow 2\pi$. Shortly after the discovery~\cite{disco}
the standard Model described theoreticaly the phenomenon~\cite{koma}: 
the bulk of 
CP violation in the kaon system is due to the mixing of CP eigenstates
$K_{1}$ and $K_{2}$, inside the physical particles \ks\ and \kl:
\begin{eqnarray}
K_{S}=K_{1}+\varepsilon K_{2}~~~~~~~~~~~~K_{L}=K_{2}+\varepsilon K_{1} \nonumber
\end{eqnarray}
This main contribution to the CP Violation is called {\it indirect} 
and it is indicated by  the parameter
$\varepsilon$. A second weaker but measurable component can arise
 in the kaon decay process~\cite{egn}:
comparing the size of CP Violation in the \kltopc\ and \kltopn\ one
measures the strength of this component, called {\it direct} and represented
by the parameter \eprim . Direct CP violation is expected to be $\sim$1/1000
of the indirect component. The ratio \eoe\ is only weakly bounded by theory
to be between 0 and 30$\times 10^{-4}$, essentially because of large 
uncertainties in the hadronic part of the computation ~\cite{theory}.
An experimental measurement of \eoe\ accurate to few 10$^{-4}$ becomes 
therefore necessary to detect direct CP-Violation.
\subsection{The observables and the NA48 method}

The measurable quantity \eoe\ is connected to the double ratio R  of
 four decay rates as follows:
\begin{eqnarray}
R=
\frac{\Gamma(K_L \rightarrow\pi^{0} \pi^{0})}{\Gamma (K_S \rightarrow \pi^{0}
 \pi^{0})}/
 \frac{\Gamma(K_L \rightarrow \pi^{+} \pi^{-})}{\Gamma (K_S \rightarrow
\pi^{+} \pi^{-})}\nonumber \\ {\approx 1 - 6 \times  \eoe ~~~~~~~~~~}\nonumber
\end{eqnarray}  
\hspace{-0.2cm}
In experimental conditions this becomes:
\begin{eqnarray}
R_{exp}=
\frac{Nb(K_L \rightarrow\pi^{0} \pi^{0})}{Nb(K_S \rightarrow \pi^{0}
 \pi^{0})}/
 \frac{Nb(K_L \rightarrow \pi^{+} \pi^{-})}{Nb(K_S \rightarrow
\pi^{+} \pi^{-})}\nonumber \\ { \approx 1 - 6 \times  \eoe _{exp}~~~~~~~~~~}\nonumber
\end{eqnarray}
where branching ratios have been replaced by the numbers of detected
events in each final mode within the acceptance of the experiment. To obtain
the true double ratio R one has to correct the measured R$_{exp}$ by 
a factor  A$_{corr}$ which would take into account all acceptance, 
trigger and  analysis effects, such that:
\begin{equation}
  R_{true}= R_{exp}+A_{corr}
\end{equation}
In practice, measuring accurately \eoe\  is equivalent to measure accurately
the correction factor A$_{corr}$. All experimental efforts will concentrate
to identify, minimise and precisely quantify all possible 
sources of biases. 
\par
Notice that since R is a double ratio, a series of effects affecting
simultaneously all four decay modes would cancel. This feature is fully 
exploited in NA48: \kltopn , \kltopc , \kstopc\ and \kstopn\ are 
concurrently recorded thanks to the use of two almost parallel
beams. This implies that beam flux variations, trigger or detector 
instabilities affect similarly at least two of the four components and 
therefore leave the double ratio invariant at first order. Geometrical
acceptance doesn't vanish in the double ratio: indeed, because of
the very different \kl\ and \ks\ lifetimes and the no-similar kinematics
of \pipic\ and \pipin\ final modes, this particular correction 
can reach up to $\pm$ 10$\%$ on R, depending on the kaon energy. In order to
avoid a such large contribution to A$_{corr}$, the \kl\ decay spectrum  
 is weighted to behave like the \ks\ one, so  that the acceptance 
correction is minimised. 
\newline
CP-conserving  processes pollute to some extent the two \kl\ samples. These
are \kethree\ and \kmuthree\ decays behaving like \ktopc\,  and 
 \ktothreepi\ with missing or fused photons and satisfying all selection
criteria of \kltopn . 
Their effect  doesn't vanish in the 
double ratio and  has  to be carefully studied.
Finaly, all residual corrections have to be evaluated and
combined into A$_{corr}$.

\section{The NA48 Beams}

\begin{figure*}
\epsfxsize30pc
%{\it $\backslash$begin\{figure$\ast$\}} 
%{\it $\backslash$end\{figure$\ast$\}}
\figurebox{16pc}{32pc}{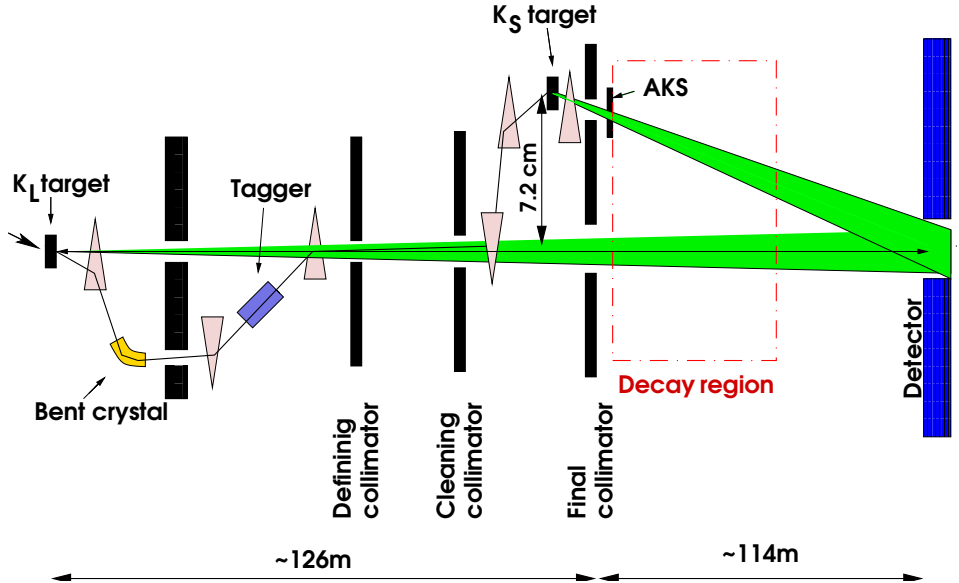}
\caption{Schematic view of the beam lines. \label{fig:beams}}
\end{figure*}

\ks\ and \kl\ beams~\cite{na48beam} are both produced by the 
450~GeV proton beam delivered by the SPS. 
An amount of $\sim$1.5$\times 10^{12}$ protons 
per pulse hit a
first berylium target located $\sim$126~m upstream of the decay volume.
The outcoming beam is let flying through long  collimation and 
bending magnets (fig.~\ref{fig:beams}). This distance of $\sim$126~m 
is enough for the \ks\ component to almost
completely decay,  so that at the exit of the final
collimator a pure \kl\ beam is dominating.

The protons surviving after the \kl\ target are deviated and sent
to cross a bent crystal~\cite{cris}. By tuning the incidence angle, one selects
the flux of transmitted protons following the crystaline planes
to be $\sim$3$\times10^{7}$ per pulse. The bending acts in an only 
6~cm length without cancelling the upstream deflection  on other
charged particles. The transmitted protons pass through a
tagging station which measures very precisely the crossing time
of all protons. After several deflections this beam is sent to strike
a second target, located $\sim$6~m before the decay volume and 
7.2~cm above the \kl\ beam. An outcoming \ks\ beam is therefore dominating
the decays at the exit of the final collimator. The two, \ks\ and \kl, 
beam axes are
slightly converging in order to intersect at the centre of the detector
120~m away and  illuminate similarly the sensitive NA48 volume.
\par
Since \ks\ and \kl\ beams are both produced by the same primary proton
beam they undergo the same intensity fluctuations, at first order.
Beam extraction from the SPS makes that \ks/\kl\ ratio can vary 
by $\pm$10$\%$
during the burst. Slower variations during the data taking periods can
originate from changes in the beam steering. Along the two years the variance
of the \ks/\kl\ intensity ratio was $\sim$9$\%$. This ratio is
continuously monitored and applied as weighting factor to the data
to make the result insensitive to eventual  unaccounted detector 
variations effects. 

Beams are tuned  such that within the decay volume the kaon energy
spectra for \ks\ and \kl\ decays are similar to $\pm$15$\%$ in the
energy range used in the analysis, namely from 70 to 170 GeV.

\section{The Detector}
The NA48 detector is designed to precisely identify and reject 
the CP-conserving
component of the \kl\ decays and to be able to register high rates
for long periods in stable conditions.
\par
Charged particles are measured by a magnetic spectrometer housed in 
a tank filled with helium. Two drift chambers 
 are located before and
two after the central dipole magnet which produce an integrated magnetic
field of 0.88~Tm. This corresponds to a transverse momentum kick of 265~MeV/c.
Each chamber~\cite{chambers} consists of eight wire planes 
oriented following four directions,
X,Y, U and V. This allows to solve reconstruction ambiguities and to 
have sufficient redundancy in case of wire inefficiencies.
The momentum resolution is $\sigma$(p)/p=0.48$\%\oplus$0.009$\times$p$\%$
where p is in GeV/c. The kaon mass is reconstructed in \pipic\ decays
with a resolution of 2.5~MeV.
\par
Neutral decays are measured by the LKr calorimeter~\cite{caloref}. 
This is a quasi-homogeneous
device consisting of $\sim$10~$m^{3}$ of liquid Krypton and of
readout electrodes, made of 50~$\mu$m$~\times$~18~mm$~\times$~125~cm  Cu-Be-Co
ribbons pulled in  longitudinal projective towers pointing to the centre
of the decay region. A cell is defined by an anode in between two cathodes, 
such that the calorimeter is segmented in $\sim$13000 cells of 
2$\times$2~cm$^{2}$ section.
The initial current readout technique reinforces the uniformity of the
energy response and provides high rate capability. Cell-to-cell
response is equalised  down to 0.15$\%$ by comparing the cluster 
energy to the momentum
measured by the spectrometer of electrons from \kethree\ decays. The pure
calorimeter resolution is found to be:
\begin{eqnarray*}
\frac{\sigma(E)}{E} = \frac{(3.2\pm0.2)\%}{\sqrt{E}}
            \oplus \frac{(9\pm1)\% }{E} \\
            \oplus       (0.42\pm0.05)\%
\end{eqnarray*}
After all corrections the energy response is linear within 0.1$\%$.
The position of a cluster is reconstructed with a resolution of 1mm
in both directions.
\par
The tagger station~\cite{tag} is made of two ladders of 
scintillator fingers, crossing
the beam  horizontally and vertically. Counter widths are decreasing
from 3.0~mm at the edges to 0.2~mm in the center of the ladder,
following the beam profile in order to equalise the counting rates. 
A small overlap between two adjacent counters guarantees the complete
coverage of the beam. A proton crossing the tagging station is measured
by at least two counters, one in the horizontal and one in the 
vertical ladder. The reconstructed time per counter is known to $\sim$140~ps
and the separation of two close protons is effective down to 4-5~ns.
\par
At the exit of the final collimator an anti-counter~\cite{akspap} vetoes all upstream 
decays of the \ks\ beam (AKS). Its sharp edge gives the geometrical reference
used for checking the stability of the energy scale.
\par 
An hadronic calorimeter following the LKr offers energy measurement 
contributing to the trigger. A series of muon counters provide time information
to identify \kmuthree\ decays.

\section{The analysis}
The data analysis for the \eoe\ measurement is done
in the following steps:
\begin{enumerate} 
\item{\pipic\ and \pipin\ decays are reconstructed}
\item{These two samples are then identified as originated from 
the \ks\ or the \kl\ beam, using  the tagging information}
\item{The remaining 3-body background is 
subtracted from the \kl\ sample in both \pipic\ and \pipin\
modes}
\item{Corrections are applied and the corresponding
 systematic uncertainties are evaluated}
\item{The double ratio is computed and its stability
with respect to several variables is checked}
\end{enumerate}
\subsection{\pipic\ and \pipin\ reconstruction}

Charged decays are reconstructed using the hits found
in the drift chambers. Hits are combined 
to tracks. Tracks found in the two chambers before the magnet
are extrapolated back and the decay vertex is defined as their
closest  distance of approach in space. The kaon energy
is computed from the opening
angle between the two tracks before the magnet and the
ratio of two pions momenta. This method makes the  
measurement independent from momentum scale uncertainties. 
Only the distance scale is sensitive to the geometry
differences between the two first chambers and also on
their relative distance, contributing to the double ratio
by (2.0$\pm$2.8)$\times 10^{-4}$.
Time information for the decay is provided by the
scintillator hodoscope with a precision of $\sim$140~ps.
\par
Neutral decays are reconstructed from energies and
positions of the photons measured by the calorimeter.
The assumption that the invariant mass of the four photons
is the kaon mass allows to
reconstruct the longitudinal coordinate of the vertex. Then combining 
the four showers to photon pairs one chooses  the two
closest to the nominal $\pi^{0}$ mass. This fully reconstructs
the \ktopn\ final modes while in case of \ktothreepi\ with missing
photons no resonant $\gamma-\gamma$ mass is found.
The calorimeter provides very precise time information
for the decay. A combination of the four photon times
gives the event time with an accuracy of $\sim$220~ps.
\par
\begin{figure} 
\epsfxsize200pt
\figurebox{1}{200pt}{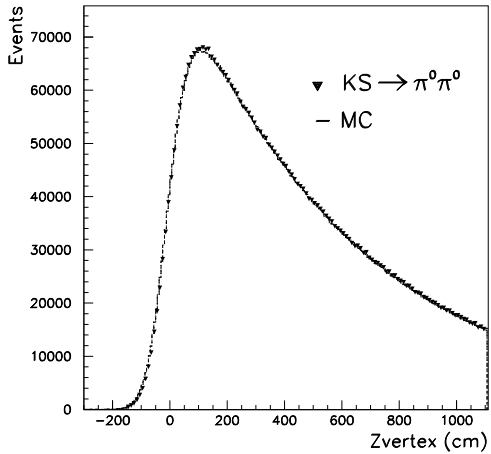}
\caption{Distribution of the reconstructed vertex position of \kstopn\ 
candidates. The nominal AKS position is at Z$_{vertex}$=0}
\label{fig:aks}
\end{figure}
The decay volume is determined such that the sensitivity
of the result on uncertainties -essentially related to
the neutral energy scale and acceptance- are minimised.
Events are accepted if the kaon energy belongs to the
range from 70 to 170~GeV and if their longitudinal decay vertex position
is no more than 3.5 times the \ks\ lifetime starting from 
the \ks\ anti-counter position. The use of a common 
decay volume implies that the energy and lifetime
definition should be the same for \pipic\ and \pipin\
decays. In charged decays the scale is given
by the geometry of the two first chambers as quoted above.
In neutral decays, the LKr information defines
both energy and vertex scales. To control the
correctness and stability of the energy scale we compare the reconstructed
position of the sharp AKS edge to its nominal value (figure~\ref{fig:aks}).
Any deviation is corrected by a global factor applied to the response of
all channels. Uncertainties on the energy reconstruction
are deeply studied and are related to energy leakage between close showers, 
non-linearities, residual energy and transverse
scales and non gaussian tails~\cite{recons}. The total effect on the
double ratio leads to  an uncertainty of $\pm$5.8$\times 10^{-4}$.

\subsection{Distinguishing a \ks\ from a \kl\ }

\begin{figure} 
\epsfxsize200pt
\figurebox{200pt}{200pt}{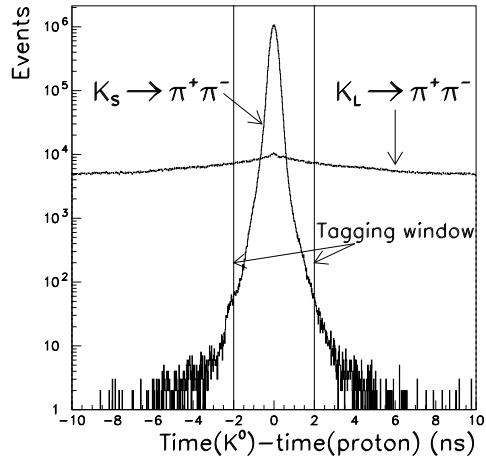}
\caption{Distribution of the event time minus the closest proton
time for \kstopc\ and \kltopc\ events identified from their vertex
position in the vertical plane}
\label{fig:tagprinc}
\end{figure}

An event will be classified as coming from the \ks\
beam if a proton is found in the tagger within $\pm$2~ns.
Figure~\ref{fig:tagprinc} shows the distribution of the event
time minus its closest proton time for \kstopc\
and \kltopc\  identified as such from their
vertex position in the vertical plane. 

In case
of true \ks\ decays, almost all events have a proton 
in coincidence within $\pm$2~ns. The small fraction
of events lying outside the tagging window
 corresponds to time reconstruction inefficiencies and it is
 called $\alpha_{SL}$. This will 
cause \ks\ decays to be mis-labelled \kl\ and it must be corrected in the 
result. On the other hand, because of the high counting
rate in the tagging station, \kl\  decays can  accidentally 
have a time coincidence with a proton. The result must therefore 
be corrected for the fraction of \kl\ decays
(called $\alpha_{LS}$) mis-accounted as \ks. In addition
the double ratio is particularly sensitive to eventual differences
of $\alpha_{LS}$ and $\alpha_{SL}$ between \twopic\ and \twopin\ 
 modes. Tagging mis-identification factors require therefore a
precise measurement in both modes.
\par
\kstopc\ decays recognised by their vertex position allow
the study of {\it{tagging inefficiency $\alpha_{SL}$}}. The fraction of 
(1.63$\pm$0.03)$\times 10^{-4}$ of events having the closest
proton time further than 2~ns (see figure~\ref{fig:tagprinc})
 have been scrutinized and
found to be  due in $\sim$80$\%$ of cases to mis-reconstruction 
of the proton time.
This is therefore symmetricaly affecting \twopic\ and
\twopin\ such that the double ratio remains invariant.
However an eventual efficiency difference of the event
time reconstruction has to be considered between the
hodoscope and the LKr calorimeter, providing information
for \twopic\ and \twopin\ events respectively. This is
studied looking at \ktopn\ and \ktothreepi\ 
events with a photon conversion. The two electron tracks
allow the time reconstruction from the hodoscope while
the photons contained in the event give the decay time
information from the calorimeter. The number of cases
where these two time estimators differ by more than
2~ns corresponds to the differential inefficiency between
the hodoscope and the calorimeter. This method allows
to show that $\alpha_{SL}$ is identical for \kstopc\
and \kstopn\ with an uncertainty of $\pm$0.5$\times 10^{-4}$.
This translates to $\pm$3$\times 10^{-4}$ uncertainty on the
double ratio.
\begin{figure}
\vspace{-0.9cm}
\epsfxsize180pt
\figurebox{}{}{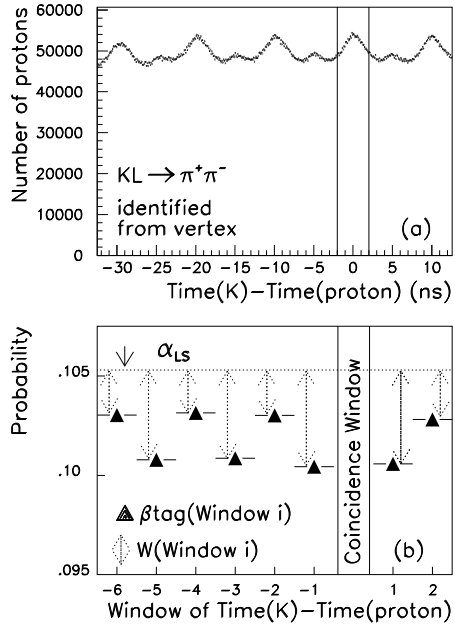}
\caption{In (a) distribution of proton times with respect to event
time for \kltopc\ events identified from their vertex. In (b) variables
$\alpha_{LS}$, $\beta_{tag}$ and W are schematically shown for 
tagged \kl\ events.}
\label{fig:tagplot}
\end{figure}
This result has been crosschecked and confirmed with \pipin\ data 
recorded with a single \ks\ beam and also with \ktopn\ events
with a subsequent Dalitz decay, where the track vertex reconstruction
in the vertical plane  provides the beam origin.
\newline
We call {\it{dilution}} the fraction $\alpha_{LS}$ of \kl\
events misidentified as \ks\ because of an accidental coincidence
with a proton. This only depends on the proton rate in the tagger and must
equally affect both decay modes. \kltopc\ events recognised by their
vertex allow to measure $\alpha_{LS}$=(10.649$\pm$0.008)$\%$ 
(see figure~\ref{fig:tagplot}.a). A direct evaluation of 
$\alpha_{LS}$ in \kltopn\ decays is not possible. We use, instead of 
the coincidence window, out-of-time windows, 4~ns wide, in 
events tagged as \kl. This is done for both \kltopc\ and 
\kltopn\ events, using  several such windows to increase the statistical
accuracy. Because \kl\ tagged events do not have any proton in 
coincidence, the dilution $\beta_{tag}$ measured in lateral windows
is smaller than  $\alpha_{LS}$ by a quantity W (see figure
~\ref{fig:tagplot}.b). W$^{+-}$ is measured
directly  in \pipic\, using \kltopc\ identified from
vertex.
For W$^{00}$ we use \ktothreepi\ events assuming that
their tagging properties are identical to those of the \kltopn\
sample. We exploited a special \kl\ run taken in 2000 to confirm
this hypothesis.
Combining all the available information we found that:
\par
~~$\Delta \alpha_{LS}=\alpha_{LS}^{00}-\alpha_{LS}^{+-}$
\par
~~~~~~~~~~~=~($\beta_{LS}^{00}+W^{00})-(\beta_{LS}^{+-}+W^{+-}$)
\par
~~~~~~~~~~=~$(4.3\pm1.8)\times 10^{-4}$
\newline
This difference indicates that \pipin\ events are recorded in slightly
higher intensity conditions than \pipic. This has been quantitatively
explained by the higher sensitivity of the charged events to the
accidental activity at both trigger and reconstruction levels. From these
two contributions one would expect  $\Delta \alpha_{LS}$ =(3.5$\pm$0.5)
$\times 10^{-4}$ in good agreement with the measurement.
$\Delta\alpha_{LS}$ implies a correction on the double ratio of
+(8.3$\pm$3.4)$\times 10^{-4}$.

\subsection{Background subtraction}

\begin{figure}
\vspace{-0.7cm}
\epsfxsize200pt
\figurebox{}{}{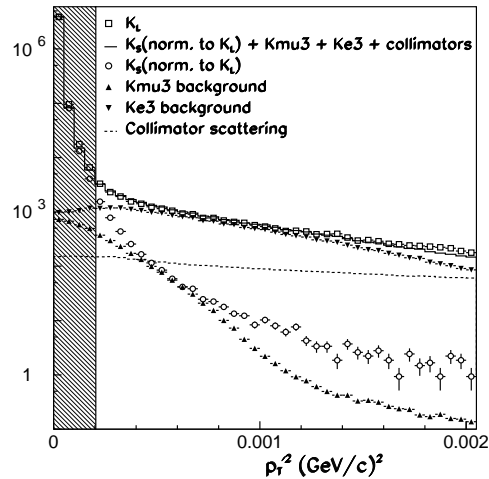}
\caption{Distribution of tranverse momentum P$_{T}^{'2}$ and
comparison of the data with all known components. The signal region is
defined for  P$_{T}^{'2} <0.0002GeV/c^{2}$}

\label{fig:chbkg}
\end{figure}

\kl\ events contain a fraction of the 3-body decays, highly
suppressed by trigger conditions and analysis cuts.
In charged mode case, \kmuthree\ and \kethree\ events can
mimic good \pipic\ decays if the electron or the muon is not
recognised by the E/p variable or the muon vetoes respectively. Because of the 
undetected neutrino of these 3-body modes the
missing transverse momentum of the decay, P$_{T}^{'2}$, is larger than  in a
\pipic\ event (see figure~\ref{fig:chbkg}). Moreover the reconstructed 
2-track invariant mass follows different distributions than in \pipic.
Studying the correlation of P$_{T}^{'2}$ versus M$_{\pi^{+}\pi^{-}}$ in 
different control regions for \pipic, \kmuthree\ et \kethree\
identified samples, allows to determine the contamination of the
signal region. 

We found that the \kltopc\ sample 
contains after all cuts 16.3$\times10^{-4}$ of 3-body residual background, dominated by
\kethree. The corresponding correction on the double ratio is
(16.8$\pm$3.0)$\times 10^{-4}$ where the error takes also into
account systematic variations related to the control region choices.
Notice that P$_{T}^{'2}$ is the component of the tranverse momentum 
which is orthogonal to the kaon line of flight, 
reconstructed in DCH1 and assumed to come from the target.
The advantage of this definition is to 
equalise the distributions for \ks\ and \kl\ decays despite
the very different positions of the corresponding targets.
Only a small asymmetry of $\le 2\times 10^{-4}$
remains between the two beams because of no gaussian tail effects.
\par
\begin{figure}
\epsfxsize200pt
\figurebox{}{}{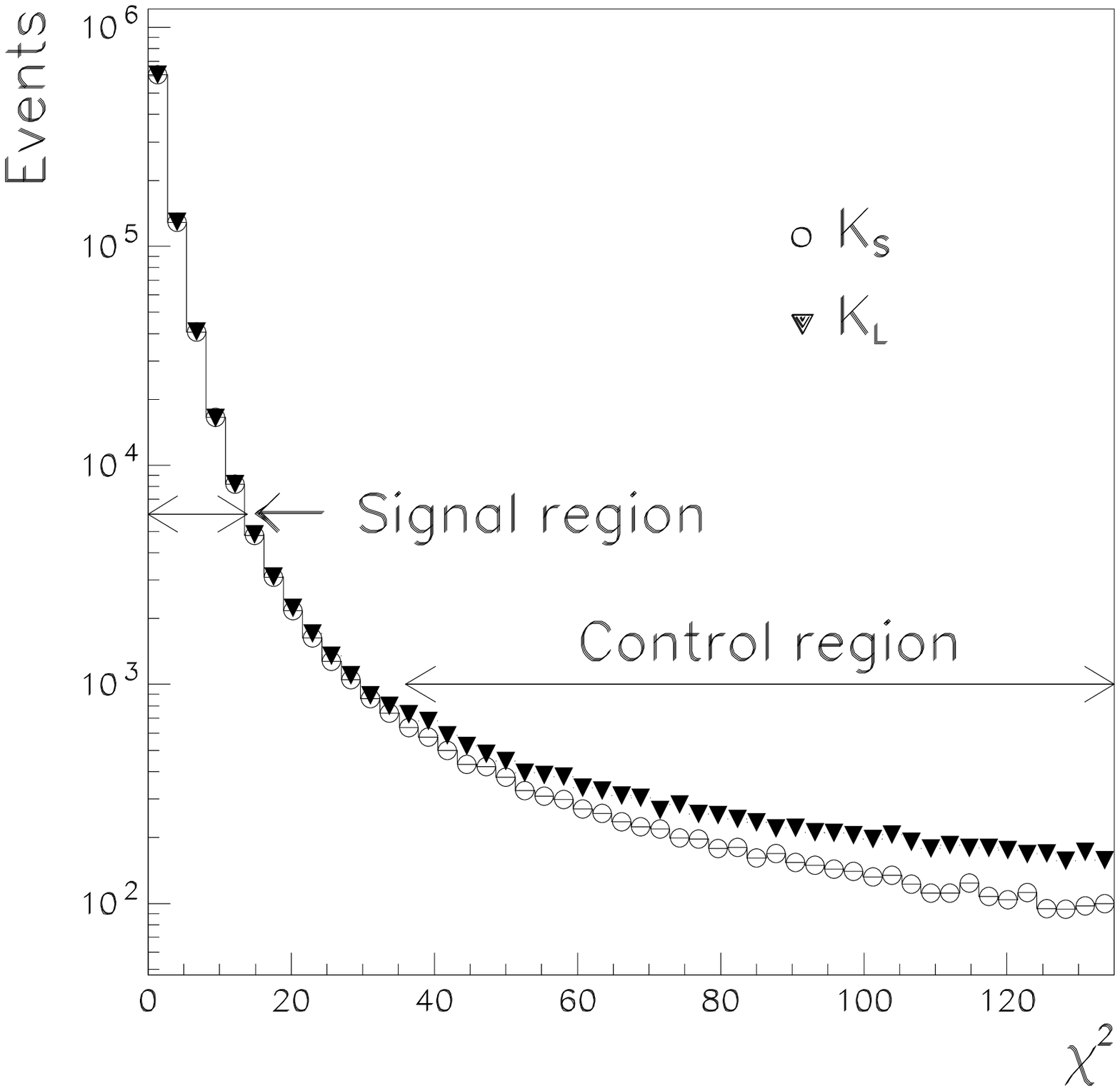}
\caption{The $\chi^{2}$ distribution for tagged \kltopn\ and \kstopn\
candidates}
\label{fig:relli}
\end{figure}
\begin{figure}
\epsfxsize220pt
\figurebox{250pt}{200pt}{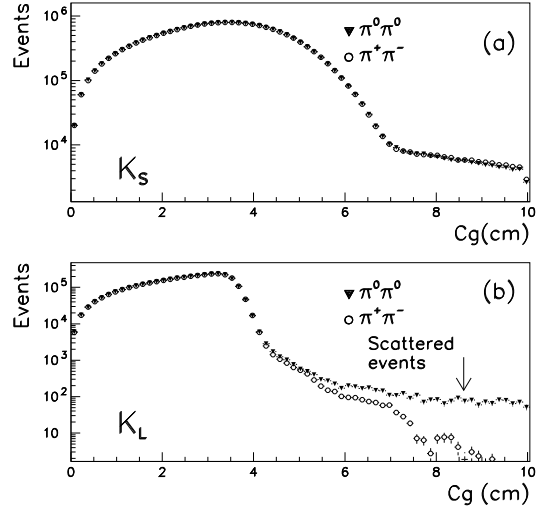}
\caption{Centre of gravity distributions for \ks\ and \kl\ events
after all cuts.}
\label{fig:cog}
\end{figure}

When \ktothreepi\ events have two photons either escaping
acceptance or fused, the two reconstructed invariant 
$\gamma-\gamma$ masses do not agree with the nominal $\pi^{0}$
one. This is observed  looking at a $\chi^{2}$
variable testing this compatibility as  shown in figure~\ref{fig:relli}. 
In \kltopn\
candidates an excess of events is observed for high values of $\chi^{2}$ 
with respect to \kstopn, where the existing tail corresponds to events 
with converted
photons or with hadronic photoproduction. The amount of remaining
background in \kltopn\ is checked in a control region, properly subtracting 
the normalised \ks\ from the \kl\ integrated population. The
extrapolated fraction of 3-body pollution under the \twopin\ signal 
results to  a correction on
the double ratio of (-5.9$\pm$2.0)$\times 10^{-4}$.
\newline
\ks\ beam is free from physical background. The $\Lambda\rightarrow p\pi^{-}$
decays are eliminated  down to $\le 10^{-4}$ level, by a track momentum asymmetry cut applied to both beams.
\newline
 Kaon scattering in the inner
edges of the collimators may modify the kinematics of an event. In  \ks\ beam, 
scattering may happen at the level at the final collimator and also when
kaons cross the anticounter. In both cases, the transverse momentum of 
the event is enhanced. After all cuts, the amount of scattered events
in the final sample is almost identical in \pipic\ and \pipin\ modes: 
the halos of centre of gravity
distributions, populated by the halo of the beam and the scattered events,
 are very similar, as shown in figure~\ref{fig:cog}.a. In \kl\ case, kaons
or accompanying neutrons in the halo of the beam can  scatter or
double-scatter in the aperture of the collimators, producing in some cases 
a \ks\ component. \ks\ particle decays
are indeed observed in \kl\ beam at high transverse momenta accumulating
around the kaon invariant mass: they all originate from the lips of the two
last collimators  and they decay with the \ks\ lifetime. These scattered 
events are asymmetrically rejected in \pipic\ and \pipin\ modes as indicated 
by the different halos of the centre of gravity distributions 
(figure~\ref{fig:cog}.b). 
The corresponding correction 
is computed to be (-9.6$\pm$2.0)$\times 10^{-4}$ on the double ratio.

\subsection{Double ratio, corrections and systematics}
After all cuts, tagging and background corrections we obtained for
the combined 98+99 data periods a statistics of 3.29M of \kltopn,
 5.21M of \kltopc\ , 14.45M of \kstopn\ and 22.22M of \kstopc decays.
Before combining these four samples 
to obtain the double ratio one has to apply some corrections.
The method of the double ratio and the simultaneous data taking of the
four modes implies that most of the effects cancel in the result
at first order.
Eventual differential effects can survive and have to
be considered. Such are for instance the tagging corrections presented
in 4.2. Another effect concerns trigger inefficiencies. In charged
mode, $\sim2\%$ of the events are lost at the trigger level essentially
because of wire inefficiencies and of information spoiled  in presence
of accidentals. This amount is identical between \ks\ and \kl\ down to
(-3.6$\pm$5.2)$\times 10^{-4}$. The neutral trigger is efficient to
99.920$\pm$0.009$\%$ level with no measurable \ks\ -\kl\ difference. 
Dead times are applied commonly in 
all four modes in order to guarantee cancellation and  identical 
intensity conditions.
\par 
The high rates of the \kl\ beam makes occasionally possible the
mismeasurement of a good event because of additional activity.
This may create losses of good events.
The high correlation of beam variations between \ks\ and \kl\
together with the concurrent data taking lead to no observable 
bias on the double ratio due to accidental losses 
within a $\pm$4.2$\times 10^{-4}$ uncertainty. 
\par

\begin{figure}
\vspace{-0.5cm}
\epsfxsize200pt
\figurebox{}{}{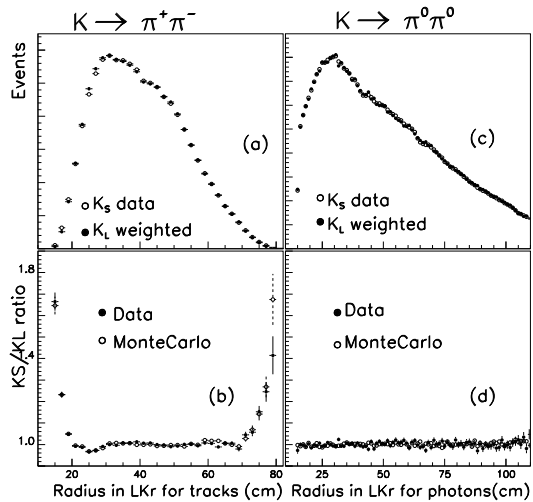}
\caption{Illumination of the calorimeter by \pipic\ tracks for
\ks\ and weighted \kl\ events in (a). The \ks/\kl\ ratio is shown in (b) where the residual effect around the
beam pipe is seen. In neutral mode \ks\ and weighted \kl\ decays
illuminate similarly the calorimeter (shown in c and d). The distributions
are shown for events in a kaon energy range from 95-100GeV}

\label{fig:mc}
\end{figure}

A no cancelling effect in the double ratio is the acceptance correction.
\ks\ and \kl\ acceptances  are not identical because of the very
different corresponding lifetimes. To avoid a correction on R as large as
$\pm$10$\%$ depending on the kaon energy, we choose to weight 
the \kl\ decay spectrum by the ratio between \ks\ and \kl\ decay 
intensities into \twopi\ taking into account the interference term 
as well. This procedure
makes by definition  the detector to be illuminated similarly by the two 
beams. A small residual  effect is due to the 0.6~mrad convergence angle 
between \ks\ and \kl\ beam lines, causing an acceptance difference 
in \pipic\ events essentially in the first chamber  around the beam pipe 
(see figure~\ref{fig:mc}). 
This is evaluated
using a large statistics Monte Carlo simulating the two beams and
detector geometry, parametrising particle interactions and using a photon
shower library generated by GEANT. The correction on the double ratio is
found to be (+26.7$\pm$4.1(stat)$\pm$4.0(syst))$\times 10^{-4}$. The
systematic error accounts for uncertainties on beam halo variations, 
beam shapes, particle interactions  in charged mode and wire inefficiencies.
\par
In table \ref{tab:corr} we give all corrections applied to the 
double ratio $R_{exp}$ and their uncertainties. The double ratio is 
corrected by A$_{corr}$=(35.9$\pm$12.6)$\times 10^{-4}$, the biggest
effects coming from acceptance, charged background,  scattering
and tagging. The quoted errors account for both statistical and
systematic uncertainties and in some cases are still statistically
limited. 

\begin{table*} 
\begin{center}
\caption{Table of corrections applied to the double ratio with
their uncertainties}\label{tab:corr}
\begin{tabular}{|l|rr|} \hline
&\multicolumn{2}{c|}{in 10$^{-4}$} \\ \hline
Trigger inefficiency in \pipic\                   &  $-3.6$       & $\pm$ 5.2  \\
AKS inefficiency                                &  $+1.1$       & $\pm$ 0.4  \\
Reconstruction effects~~~ \begin{tabular}{@{}l} of \pipin\ \\ of \pipic\ \end{tabular} &
\begin{tabular}{r@{}}    ---   \\  $+2.0$  \end{tabular} &
\begin{tabular}{r@{}}   $\pm$ 5.8 \\  $\pm$ 2.8  \end{tabular} \\
Background subtraction \begin{tabular}{@{}l} to \pipin\ \\ to \pipic\ \end{tabular} &
\begin{tabular}{r@{}} $-5.9$ \\ $+16.9$ \end{tabular} &
\begin{tabular}{r@{}} $\pm$ 2.0  \\ $\pm$ 3.0  \end{tabular} \\
Beam scattering in \kl\              &  $-9.6$       & $\pm$ 2.0  \\
Accidental tagging                              &  $+8.3$       & $\pm$ 3.4  \\
Tagging inefficiency                            &  ---          & $\pm$ 3.0  \\
Acceptance correction \begin{tabular}{@{}l} statistical \\ systematic \end{tabular}
                              & $+26.7$ &
     \begin{tabular}{r@{}} $\pm$ 4.1 \\ $\pm$ 4.0 \end{tabular}  \\
Accidental activity                             &  ---          & $\pm$ 4.4  \\
Long term intensity variations of \ks/\kl\  &  ---   & $\pm$ 0.6  \\ \hline
Total A$_{corr}$                            & $+35.9$       & $\pm$ 12.6 \\
\hline
\end{tabular}
\end{center}
\end{table*}

\begin{figure*}
\epsfxsize25pc
\figurebox{25pc}{37pc}{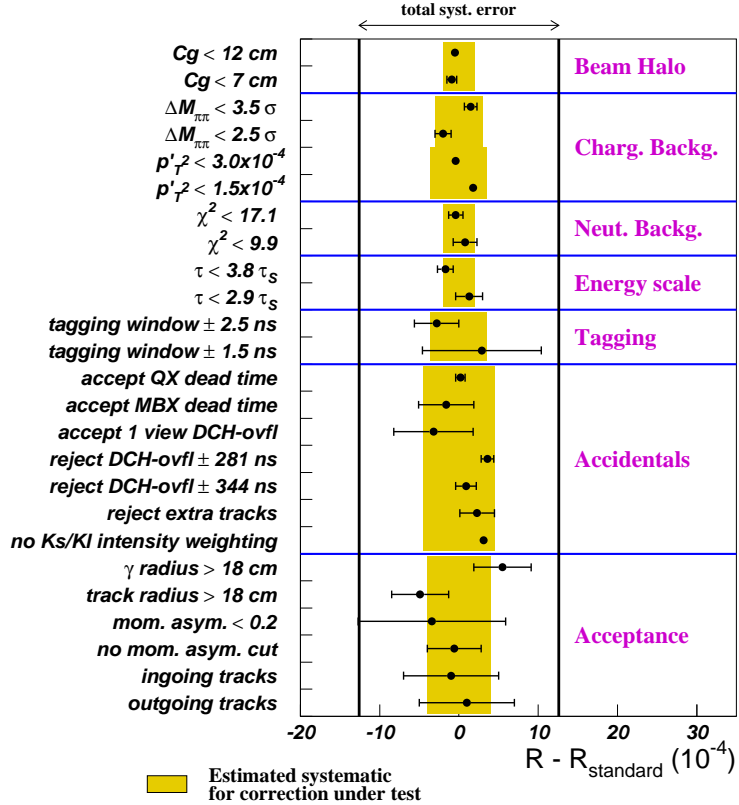}
\caption{Stability of the double ratio when varying selection
criteria}

\label{fig:syst}
\end{figure*}
The double ratio is computed in 20 bins of 5~GeV in kaon energy 
in order to further
decrease the effect of the remaining energy spectra difference between
\kl\ and \ks. All corrections are applied in bins and the 20 values
are averaged using an unbiased estimator.
The result is:
\vspace{0.3cm}
\begin{center}
R=0.99098$\pm$0.00101(stat)$\pm$0.00126(syst)
\end{center}
\vspace{0.3cm}
A large number of systematic checks were devoted to verify the
stability of the result with a change in the event selection. We cumulate
the output of the most important tests in figure~\ref{fig:syst} where 
the deviations of 
the double ratio with respect to  its standard value are shown for a series
of modifications we did  in the procedure. All these checks 
demonstrated the stability of the result within the allowed systematic
range. Moreover, the result was found to be stable with the 
data taking time , time within the burst, proton revolution phase in the SPS,
magnetic field sign in the spectrometer and 50~Hz phase.

The corresponding \eoe\ value on 98+99 data~\cite{newresult} is obtained 
subtracting R from 1 and dividing by six:
\vspace{0.3cm}
\begin{center}
\eoe=(15.0$\pm$1.7(stat)$\pm$2.1(syst))$\times 10^{-4}$
\end{center}
\vspace{0.3cm}
which, combined with the published 97 result gives: 
\begin{center}
\eoe=(15.3$\pm$2.6)$\times 10^{-4}$
\end{center}
\vspace{0.3cm}
This result confirms the existence of a direct CP-Violating
component in the neutral kaon decays at the level of 5.9$\sigma$.
The 2001 new world average, combining the four most precise experimental
results from NA31~\cite{na31}, E731~\cite{e731}, KTeV~\cite{ktev} and 
the combined 97+99+99 NA48 value becomes:
\vspace{0.3cm}
\begin{center}
\eoe=(17.2$\pm$1.8)$\times 10^{-4}$
\end{center}
\section{Review of the NA48 rare decay program}

Several \ks\ and \kl\ rare decay channels have been looked at
 in NA48. Table \ref{tab:rare} gives a non exhaustive list of observations
or limits obtained with NA48 data. We will concentrate here on two results 
concerning the recently updated analysis of
$K_{L}\rightarrow\pi^{0}\gamma\gamma$ and the CP-Violating mode 
$K_{L}\rightarrow\pi^{+}\pi^{-}$e$^{+}$e$^{-}$. More information on rare
decays in NA48 can be found in references~\cite{rareresults}.
\begin{table*}
\begin{center}
\caption{List of some of the rare decays results obtained in NA48}
\label{tab:rare}
\begin{tabular}{|l|r|r|r|} \hline
Mode  & Branching ratio & Events & Status \\
\hline
$K_{L}\rightarrow \pi^{0}\gamma\gamma$ & (1.36$\pm$0.05)$\times 10^{-6}$&2588 &preliminary\\
\hline
$K_{S}\rightarrow \pi^{0}$ e$^{+}$e$^{-}$ & $< 1.4\times 10^{-7}$ (90$\%$~CL) & &published\\
\hline
$K_{S}\rightarrow \gamma\gamma$ & (2.58$\pm$0.42)$\times 10^{-6}$ & 149 &published \\
\hline
$K_{S}\rightarrow\pi^{+}\pi^{-}$e$^{+}$e$^{-}$ & (4.3$\pm$0.4)$\times 10^{-5}$ 
& 921 &preliminary\\
\hline
$K_{L}\rightarrow\pi^{+}\pi^{-}$e$^{+}$e$^{-}$ & (3.1$\pm$0.2)$\times 10^{-7}$ 
& 1337 &preliminary\\
\hline
$K_{L}\rightarrow$e$^{+}$e$^{-}$e$^{+}$e$^{-}$ & (3.7$\pm$0.4)$\times 10^{-8}$ & 132&preliminary\\
\hline
$K_{L}\rightarrow$e$^{+}$e$^{-}\gamma $ & (1.06$\pm$0.05)$\times 10^{-5}$ &
6864& published\\
\hline
$K_{L}\rightarrow$e$^{+}$e$^{-}\gamma\gamma$ & (6.3$\pm$0.5)$\times 10^{-7}$ &
492 &preliminary\\
\hline
\end{tabular}
\end{center}
\end{table*}

\subsection{$K_{L}\rightarrow\pi^{0}\gamma\gamma$}
The interest of this mode resides in the fact that, in Chiral Perturbation
Theory one needs to include O(p$^{6}$) calculations and vector meson
contribution to reproduce the observed rate. These two terms account
actually for 1/3 of the total amplitude. Moreover the measurement of
this decay gives constraints to the CP conserving amplitude contributing
to $K_{L}\rightarrow\pi^{0}$e$^{+}$e$^{-}$, via a two photon intermediate 
state.
\newline
The reconstruction of this mode uses information of the LKr. An event must
contain one $\gamma-\gamma$ pair compatible with $\pi^{0}$ mass and a
second $\gamma-\gamma$ pair not compatible. The main
background comes from \kltopn\ and \ktothreepi\ decays. \kltopn\ with
a photon conversion before the magnet destroys the $\pi^{0}$ reconstruction.
\begin{figure}
\vspace{-0.6cm}
\epsfxsize200pt
\figurebox{}{}{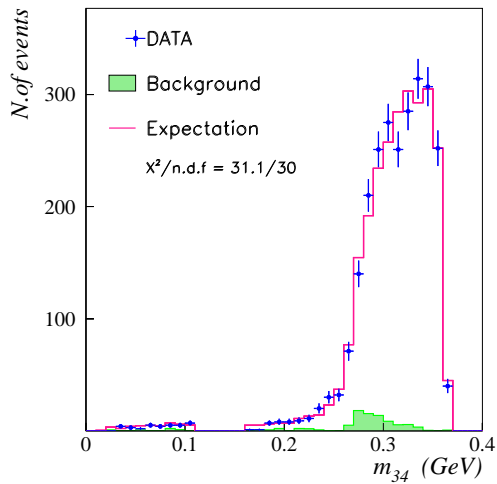}
\caption{Distribution of the invariant $\gamma-\gamma$ mass, M$_{34}$, 
of the non resonant photons}
\label{fig:klpgg}
\end{figure}
These cases are rejected requiring no hits in the first three drift chambers.
\ktothreepi\ decaying inside the collimator can fake the signal. Dedicated
variables have been studied to reduce this background to a small level.
In total we have a signal of 2588 events found in the 98+99 data sample, 
with an estimated background  of only 3.3$\%$. 
The distribution of the invariant $\gamma-\gamma$ mass for the non resonant
photon pair is shown in figure~\ref{fig:klpgg}. The low mass region 
(M$_{\gamma\gamma}<$ 260~MeV) is
particularly sensitive to the amount of vector meson production, represented
by the coupling constant a$_{V}$. 
Fitting
our data with a likelihood function including the shape predicted by 
$\chi$PT calculations up to O(p$^{6}$) and for several 
a$_{V}$ values, we found the best $\chi^{2}$ for:
\begin{center} 
\vspace{0.3cm}
a$_{V}$=-0.46$\pm$0.03(stat)$\pm$0.04(syst+theory)
\end{center}
\vspace{0.3cm}
Using this coupling constant we obtain the following branching ratio:
\vspace{0.3cm}
\newline
BR($K_{L}\rightarrow\pi^{0}\gamma\gamma$)=\\
\hspace{1cm}
(1.36$\pm$0.03(stat)$\pm$0.4(syst))$\times 10^{-6}$ \\
\newline
where the systematic error is shared between  experimental
and theoretical uncertainties. A publication is in preparation.

\subsection{$K_{L}\rightarrow\pi^{+}\pi^{-}$e$^{+}$e$^{-}$ }
CP Violation appears in the kaon system through small quantities,
namely \epsi$\sim$0.2$\%$ and \eprim$\sim 10^{-6}$.
The $K_{L}\rightarrow\pi^{+}\pi^{-}$e$^{+}$e$^{-}$ decay demonstrates
that \kl\ can also exhibit strong CP-violating phenomena. Indeed, this
channel arises through two diagrams
essentially: the inner-breamsstrahlung, violating CP, and
the direct emission, CP conserving process. Because the
ratio of these two contributions is almost one 
($\sim \eta\times$(2M$_{K}$/E$_{\gamma}$)$^{2}$ where E$_{\gamma}$ is the
virtual photon energy in the centre of mass), the
interference is enhanced. Looking at the angle $\phi$ between the
\pipic\ and e$^{+}$e$^{-}$ planes one can observe strong CP asymmetries.
\begin{figure}
\vspace{-0.6cm}
\epsfxsize200pt
\figurebox{}{}{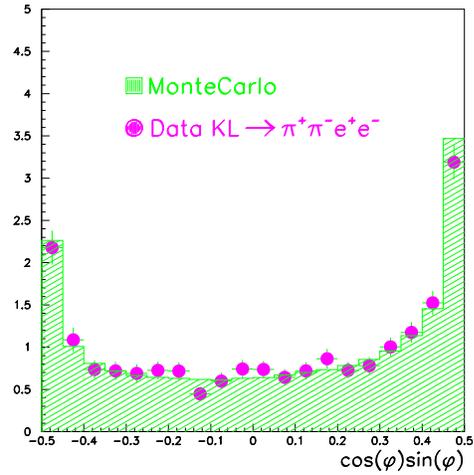}
\caption{Cos($\phi$)sin($\phi$) distribution for $K_{L}\rightarrow\pi^{+}\pi^{-}$e$^{+}$e$^{-}$ decays}
\label{fig:pipiee}
\end{figure}

We detected 1337 events after all cuts. 
The cos($\phi$)sin($\phi$)
distribution is shown in figure~\ref{fig:pipiee} where it is also compared
with Monte Carlo simulation. After acceptance corrections the measured
asymmetry is:
\begin{center}
\vspace{0.3cm}
A=(13.9$\pm$2.7(stat)$\pm$2.0(syst))$\%$
\end{center}
\vspace{0.3cm}
The branching ratio of this decay is found to be:
\vspace{0.3cm}
\begin{center}
BR=(3.1$\pm$0.1(stat)$\pm$0.2(syst))$\times 10^{-7}$
\end{center}
\vspace{0.3cm}
The measured asymmetry is in agreement with theoretical predictions 
and previous results~\cite{ppee}.
Moreover NA48 detected the first signal of  $K_{s}\rightarrow\pi^{+}\pi^{-}$e$^{+}$e$^{-}$ decays~\cite{ksppee}. 
This mode conserves CP symmetry. It allows therefore  
to check eventual effects of final states interactions  on the  
cos($\phi$)sin($\phi$) distribution, in 
which case, the \kl\ mode would have been also affected. In \ks\ beam
we measured an asymmetry compatible with zero, based in a sample of 
921 events. 
The observed asymmetry of 13.9$\%$ in 
$K_{L}\rightarrow\pi^{+}\pi^{-}$e$^{+}$e$^{-}$ decays is  therefore only 
due to
the indirect CP Violation present in this rare process. 

\section*{Acknowledgments}
It's a pleasure to thank Cecilia Voena, the scientific secretary of the
session,  for her help. 
It's amazing how
well and happily  almost 800 people have lived, worked and 
have had  nice time together in Roma. 
Thank you profoundly, Juliet, Paolo and the Committee, for the organisation 
of this excellent conference and please, do it again.

\end{document}